\documentclass[useAMS,usenatbib,usegraphicx]{mn2e}
\newcommand{\beq}[1]{\begin{equation}\label{#1}}
 \newcommand{\eeq}{\end{equation}}
 \newcommand{\bea}{\begin{eqnarray}}
 \newcommand{\eea}{\end{eqnarray}}
\def\la{\mathrel{\hbox{\rlap{\hbox{\lower4pt\hbox{$\sim$}}}\hbox{$<$}}}}

\usepackage{epsf}
\usepackage{amssymb,amsmath}

\title[Testing IDE with $H(z)$]
{Testing the phenomenological interacting dark energy with observational $H(z)$ data
}

\author[Cao, Liang and Zhu]
{Shuo Cao, Nan Liang \thanks{liangn@bnu.edu.cn}and Zong-Hong Zhu \thanks{zhuzh@bnu.edu.cn}\\
$^1$Department of Astronomy, Beijing Normal University, 100875,
Beijing, China}

\begin{document}

\date{\today}

\voffset- .5in

\pagerange{\pageref{firstpage}--\pageref{lastpage}} \pubyear{}

\maketitle

\label{firstpage}

\begin{abstract}
In order to test the possible interaction between dark energy and
dark matter, we investigate observational constraints on a
phenomenological scenario, in which the ratio between the dark
energy and matter densities is proportional to the power law case of
the scale factor, $r\equiv (\rho_X/\rho_m)\propto a^{\xi}$. By using
the Markov chain Monte Carlo method, we constrain the
phenomenological interacting dark energy model with the newly
revised $H(z)$ data, as well as the cosmic microwave background
(CMB) observation from the 7-year Wilkinson Microwave Anisotropy
Probe (WMAP7) results, the baryonic acoustic oscillation (BAO)
observation from the spectroscopic Sloan Digital Sky Survey (SDSS)
data release 7 (DR7) galaxy sample and the type Ia supernovae (SNe
Ia) from Union2 set. The best-fit values of the model parameters are
$\Omega_{m0}=0.27_{-0.02}^{+0.02}(1\sigma)_{-0.03}^{+0.04}(2\sigma)$,
$\xi=3.15_{-0.50}^{+0.48}(1\sigma)_{-0.71}^{+0.72}(2\sigma)$, and
$w_X=-1.05_{-0.14}^{+0.15}(1\sigma)_{-0.21}^{+0.21}(2\sigma)$, which
are more stringent than previous results. These results show that
the standard $\Lambda$CDM model without any interaction remains a
good fit to the recent observational data; however, the interaction
that the energy transferring from dark matter to dark energy is
slightly favored over the interaction from dark energy to dark
matter. It is also shown that the $H(z)$ data can give more
stringent constraints on the phenomenological interacting scenario
when combined to CMB and BAO observations, and the confidence
regions of $H(z)$+BAO+CMB, SNe+BAO+CMB, and $H(z)$+SNe+BAO+CMB
combinations are consistent with each other.

\end{abstract}


\section{Introduction}\label{sec1}
The fact that the universe is undergoing an accelerating expansion
has been supported and comfirmed by many cosmological observations,
such as the luminosity distances of Type Ia Supernovae [SNe Ia,
\citep{Riess98,Perlmutter99,Astier06,Hicken09,Amanullah10}], cosmic
microwave background (CMB) from Wilkinson Microwave Anisotropy Probe
[WMAP, \citep{Spergel03,Spergel07,Komatsu09,Komatsu10}], and the
large scale structure from Sloan Digital Sky Survey [SDSS,
\citep{Tegmark04,Eisenstein05}]. In order to explain this mysterious
phenomenon, the existence of dark energy with negative pressure,
which dominates the total energy density and causes an accelerating
expansion of our universe at late times, has been widely proposed.
The most simple candidate of dark energy models is considered to be
in the form of vacuum energy density or cosmological constant
($\Lambda$), with a equation of state (EoS):
$w_\Lambda=p_\Lambda/\rho_\Lambda\equiv-1$. However, the
corresponding $\Lambda$CDM model is always entangled with the
coincidence problem: The density of the cosmic component decreases
with $\rho_i \propto a^{-3(1+w_i)}$ during the expansion of our
universe; therefore the matter density ($\rho_m$) decreases with
$a^{-3}$, and the cosmological constant density ($\rho_\Lambda$) do
not change in the cosmic expansion; however, the dark energy density
is comparable with the dark matter density today.

In order to alleviate the coincidence problem, many alternative
models include the scalar field models with dynamical EoS (e.g., the
quintessence \citep{Ratra88,Caldwell98}, phantom
\citep{Caldwell02,Caldwell03}, k-essence
\citep{Armendariz-Picon01,Chiba02}, as well as quintom model
\citep{Feng05} have been proposed, however, the nature of dark
energy is still unknown. It is natural to consider the possibility
of exchanging energy between dark energy and dark matter. In the
interacting scenario, $\rho_m$ could decrease slower than $a^{-3}$
during the cosmic expansion to alleviate the coincidence problem.
Considering that the dark energy ($\rho_X$, assuming a constant EoS,
$w_X\equiv\textit{const}$) and the dust matter including the baryon
and dark matter component ($\rho_m=\rho_b+\rho_{DM}$) exchange
energy through an interaction term $Q$,
 \bea
 &&\dot{\rho}_X+3H\rho_X(1+w_X)=-Q,\nonumber\\
 &&\dot{\rho}_m+3H\rho_m=Q,
 \eea
which preserves the total energy conservation equation
$\dot{\rho}_{tot}+3H\left(\rho_{tot}+p_{tot}\right)=0$, where
$\rho_{tot}=\rho_{X}+\rho_{m}$. Various interaction theoretical
models have been put forward and studied
\citep{Amendola00,Zimdahl01,Chimento03,Guo05,GuoZhang05,Wei06,Wei07a,
Wei07b,Zhang09,CGS09,Baldi10,Cervantes-Cota10,Zhang10,Cai10}, but
the format of interaction term ¡°$Q$¡± still can not be determined
from fundamental physics.

On the other hand, the interacting dark energy can be investigated
in a phenomenological way with minimal underlying theoretical
assumptions. \citet{Dalal01} proposed a simple phenomenological
scenario in which the ratio between the dark energy and matter
densities is proportional to the power law case of the scale factor,
\bea r\equiv \frac{\rho_X}{\rho_m}\propto a^{\xi}, \eea where $\xi$
is a key parameter, which quantifies the severity of the coincidence
problem. The special cases $\xi=3$ and $\xi=0$ correspond to the
$\Lambda$CDM model and the self-similar solution without coincidence
problem respectively. Hence, any solution with a scaling parameter
$0<\xi<3$ makes the coincidence problem less severe \citep{Pavon04}.
Considering a flat FRW universe with $\Omega_{X0}+\Omega_{m0}=1$
(where $\Omega_{X0}$, $\Omega_{m0}$ are the present value of density
parameter of the dark energy and dust matter, respectively), and
setting
$r=a^{\xi}\Omega_{X0}/\Omega_{m0}$, 
we can obtain the corresponding interaction term
\citep{Guo07,Wei07b}
\begin{equation}
Q=-H\rho_m(\xi+3w_X) \Omega_{X},
 \end{equation}
where
$\Omega_{X}={(1-\Omega_{m0})}/{[1-\Omega_{m0}+\Omega_{m0}(1+z)^\xi]}$.
In this case, the standard cosmology without interaction between the
dark sector is characterized by $\xi+3\omega_X=0$, while
$\xi+3\omega_X\neq 0$ denotes non-standard cosmology. The case
$\xi+3\omega_X<0$ denotes that the energy is transferred from dark
energy to dark matter ($Q>0$), which can alleviate  the coincidence
problem; whereas the case $\xi+3\omega_X>0$ indicates that the
energy is transferred from dark matter to dark energy ($Q<0$), in
which the coincidence problem is more severe.

The Friedmann equation of the phenomenological scenario with a
constant EoS of dark energy can be expressed as
\begin{eqnarray}
 E(z)^2&\equiv&H^2/H_0^2 \nonumber\\
&=& (1+z)^{3}\left[\Omega_{m0}+
(1-\Omega_{m0})(1+z)^{-\xi}\right]^{-3w_X/\xi}.
\end{eqnarray}
From Eq. (4), the simple phenomenological interacting scenario has
been constrained from many cosmological observations. \citet{Guo07}
considered the cosmological constraints on this phenomenological
scenario with the luminosity distances ($d_L$) of SNe Ia data from
the Supernova Legacy Survey [SNLS, \citep{Astier06}], the shift
parameter $R$ from the 3-year WMAP [WMAP3, \citep{Spergel07}]
results, and the distance parameter $A$ of the baryon acoustic
oscillation [BAO, \citep{Eisenstein05}] observation. Recently,
\citet{Chen10} tested this phenomenological form by combining the
397 SNe Ia from the Constitution Set \citep{Hicken09}, the  $R$
parameter from the 5-year WMAP [WMAP5, \citep{Komatsu09}] results
and $A$ parameter, to show that the $\Lambda$CDM model still remains
a good fit to the recent observational data, as well as, the
coincidence problem indeed exists and is quite severe. More
recently, \citet{Wei10a} constrained the general type which is
characterized by $\rho_{_X}/\rho_m=f(a)$ (where $f(a)$ can be any
function of scale factor $a$) from the latest observational data
including the 397 SNIa data set, the 7-year WMAP [WMAP7,
\citep{Komatsu10}] results and $A$ parameter. Some relevant works in
other phenomenological way can be found in
\citet{Wang05,Wei10b,Costa10}.

For cosmological observations, it is well known that SNe Ia, CMB and
BAO use the distance scale (e.g., $d_L$, $R$, or $A$) measurement to
determine the cosmological parameters. However, we need to integrate
the Hubble parameter to get the distance scale. The integral cannot
take the fine structure of $H(z)$ into consideration and lose some
important information compiled in it. Therefore, it is more
rewarding to investigate the observational $H(z)$ data directly
\citep{Cao11}.

The observational Hubble parameter depends on the differential age
as a function of redshift $z$ in the form
 \begin{equation}
 H(z)=-\frac{1}{1+z}\frac{dz}{dt}\,.
 \end{equation}
\citet{Jimenez03} demonstrated the feasibility of the method by
applying it to a $z\sim 0$ sample. \cite{hz1} determined nine $H(z)$
data in the range $0\leq z \leq 1.8$ by using the differential ages
of passively evolving galaxies determined from the Gemini Deep Deep
Survey [GDDS, \citep{Abraham04}] and archival data
\citep{Treu01,Treu02,Nolan03a,Nolan03b}. \citet{Wei07a,Wei07b} used
the nine observational $H(z)$ data to constrain the interacting dark
energy models, and some other relevant works useing these $H(z)$
data for cosmological constraint include
\citet{Samushia06,Lazkoz07,Yi07,Wu07a,Wu07b,Zhang07,Kurek08,Sen08,Zhang08,Xu08,Lin09,Zhai10}.
Recently, \citet{hz2} revised the $H(z)$ data at 11 different
redshifts from the differential ages of red-envelope galaxies. On
the other hand, \citet{hz3} determined other two Hubble parameter
data at $z=0.24$ and $z=0.43$ from observations of BAO peaks.
\citet{Cao11} used these newly observational $H(z)$ data to
constrain on the Interacting Dark Matter (IDM) scenario, and some
other works for cosmological constraint can be found in
\citet{Wang09,Gong10,Liang10,Liang11b,Xu10,Ma11}. For recent review
of the observational $H(z)$ data, 
see e.g. \citet{ZML10}.

In this paper, we investigate observational constraints on the
simple phenomenological interacting scenario by performing a Markov
Chain Monte Carlo (MCMC) analysis. For cosmological observations, we
focus on the newly $H(z)$ data from the differential ages of
red-envelope galaxies \citep{hz2} and observations of BAO peaks
\citep{hz3}.  In order to break the degeneracy of model parameters,
we combine the $H(z)$ data with the CMB observation from the WMAP7
results \citep{Komatsu10} and the  BAO distance ratio ($d_z$) from
the spectroscopic Sloan Digital Sky Survey (SDSS) data release 7
(DR7) galaxy sample\citep{Percival10}. For examining the role of the
$H(z)$ data played in cosmological constraints, we also add the
newly revised Union2 set which consists of 557 SNe Ia
\citep{Amanullah10}. This paper is organized as follows: In
section~\ref{sec2}, we introduce the observational data including
$H(z)$, BAO, CMB, as well as SNe Ia. In section~\ref{sec3}, we
perform a Markov Chain Monte Carlo analysis spanning the full
parameter space of the model using different data sets to constrain
the phenomenological interacting model. Finally, we summarize the
main conclusions in Section~\ref{sec4}.

\section{Observational data}\label{sec2}
In order to break the degeneracy of model parameters, we combine the
$H(z)$ data with the CMB observation from the WMAP7 results
\citep{Komatsu10} and the  BAO observation from the SDSS DR7 galaxy
sample\citep{Percival10}. We also add the newly revised Union2 set
which consists of 557 SNe Ia \citep{Amanullah10} to examine the role
of the $H(z)$ data played in cosmological constraints.


For the observational $H(z)$ data, we adopt the 11  data obtained
from the differential ages of red-envelope galaxies \citep{hz2}, and
two data at $H(z=0.24)=76.69\pm3.61\rm{Mpc}^{-1}$, and
$H(z=0.43)=86.45\pm4.96\rm{Mpc}^{-1}$ determined from observations
of BAO peaks \citep{hz3}.
The $\chi^2$ value of the $H(z)$ data can be given by
\begin{equation}
\label{chi2H}
\chi^2_H=\sum_{i=1}^{13}\frac{[H(z_i)-H_{obs}(z_i)]^2}{\sigma_{h,i}^2},
\end{equation}
where $\sigma_{h,i}$ is the $1\sigma$ uncertainty in the $H(z)$
data.

%

For the CMB observation, we use the data set including  the acoustic
scale ($l_a$), the shift parameter ($R$), and the redshift of
recombination ($z_{\ast}$). From the WMAP7 measurement, the best-fit
values of the data set are \citep{Komatsu10}
\begin{eqnarray}
\hspace{-.5cm}\bar{\textbf{P}}_{\rm{CMB}} &=& \left(\begin{array}{c}
{\bar l_a} \\
{\bar R}\\
{\bar z_{\ast}}\end{array}
  \right)=
  \left(\begin{array}{c}
302.09 \pm 0.76\\
1.725\pm 0.018\\
1091.3 \pm 0.91 \end{array}
  \right).
 \end{eqnarray}
The $\chi^2$ value of the CMB observation can be expressed as
\citep{Komatsu10}
\begin{eqnarray}
\chi^2_{\mathrm{CMB}}=\Delta
\textbf{P}_{\mathrm{CMB}}^\mathrm{T}{\bf
C_{\mathrm{CMB}}}^{-1}\Delta\textbf{P}_{\mathrm{CMB}},
\end{eqnarray}
where $\Delta\bf{P_{\mathrm{CMB}}} =
\bf{P_{\mathrm{CMB}}}-\bf{\bar{P}_{\mathrm{CMB}}}$, and the
corresponding inverse covariance matrix is
\begin{eqnarray}
\hspace{-.5cm} {\bf C_{\mathrm{CMB}}}^{-1}=\left(
\begin{array}{ccc}
2.305 &29.698 &-1.333\\
29.698 &6825.270 &-113.180\\
-1.333 &-113.180 &3.414
\end{array}
\right).
\end{eqnarray}
The acoustic scale can be expressed as
\begin{equation}
l_a=\pi\frac{\Omega_\mathrm{k}^{-1/2}sinn[\Omega_\mathrm{k}^{1/2}\int_0^{z_{\ast}}\frac{dz}{E(z)}]/H_0}{r_s(z_{\ast})},
\end{equation}
where $r_s(z_{\ast})
={H_0}^{-1}\int_{z_{\ast}}^{\infty}c_s(z)/E(z)dz$ is the comoving
sound horizon at photo-decoupling epoch. The shift parameters can be
expressed as
\begin{equation}
R=\Omega_{\mathrm{m0}}^{1/2}\Omega_\mathrm{k}^{-1/2}sinn\bigg[\Omega_\mathrm{k}^{1/2}\int_0^{z_{\ast}}\frac{dz}{E(z)}\bigg].
\end{equation}
The redshift of recombination  is
$z_{\ast}=1048[1+0.00124(\Omega_bh^2)^{-0.738}(1+g_{1}(\Omega_{\mathrm{m,0}}h^2)^{g_2})]$,
where
$g_1=0.0783(\Omega_bh^2)^{-0.238}(1+39.5(\Omega_bh^2)^{-0.763})^{-1}$
and $g_2=0.560(1+21.1(\Omega_bh^2)^{1.81})^{-1}$ \citep{Hu96}.

For the BAO observation, we use the measurement of the BAO distance
ratio ($d_z$) at $z=0.2$ and $z=0.35$ \citep{Percival10}. From SDSS
data release 7 (DR7) galaxy sample, the best-fit values of the data
set ($d_{0.2}$, $d_{0.35}$) are \citep{Percival10}
\begin{eqnarray}
\hspace{-.5cm}\bar{\bf{P}}_{\rm{BAO}} &=& \left(\begin{array}{c}
{\bar d_{0.2}} \\
{\bar d_{0.35}}\\
\end{array}
  \right)=
  \left(\begin{array}{c}
0.1905\pm0.0061\\
0.1097\pm0.0036\\
\end{array}
  \right).
 \end{eqnarray}
The $\chi^2$ value of the BAO observation from SDSS DR7 can be
expressed as \citep{Percival10}
\begin{eqnarray}
\chi^2_{\mathrm{BAO}}=\Delta
\textbf{P}_{\mathrm{BAO}}^\mathrm{T}{\bf
C_{\mathrm{BAO}}}^{-1}\Delta\textbf{P}_{\mathrm{BAO}},
\end{eqnarray}
where  the corresponding inverse  covariance matrix is
\begin{eqnarray}
\hspace{-.5cm} {\bf C_{\mathrm{BAO}}}^{-1}=\left(
\begin{array}{ccc}
30124& -17227\\
-17227& 86977\\
\end{array}
\right).
\end{eqnarray}
The BAO distance ratio  can be expressed as
\begin{equation}
d_z=\frac{r_s(z_d)}{D_V(z_{\mathrm{BAO}})},
\end{equation}
where the distance scale $D_V$ is given by \citep{Eisenstein05}
\begin{equation} D_V(z_{\mathrm{BAO}})=\frac{1}{H_0}\big
[\frac{z_{\mathrm{BAO}}}{E(z_{\mathrm{BAO}})}\big(\int_0^{z_{\mathrm{BAO}}}\frac{dz}{E(z)}\big
)^2\big]^{1/3}~,
\end{equation}
and $r_s(z_d)$ is the comoving sound horizon at  the drag epoch,
where
$z_{d}=\{{1291(\Omega_{\mathrm{m0}}h^2)^{0.251}}/{[1+0.659(\Omega_\mathrm{m0}h^2)^{0.828}]}\}[(1+b_{1}(\Omega_{b}h^2)^{b_2})]$,
where
$b_1=0.313(\Omega_{\mathrm{m,0}}h^2)^{-0.419}[1+0.607(\Omega_{\mathrm{m,0}}h^2)^{0.674}]^{-1}$
and
$b_2=0.238(\Omega_{\mathrm{m,0}}h^2)^{0.223}$\citep{Eisenstein98}.


SNe Ia provide the most direct indication of the accelerated
expansion of the universe. We add SNe Ia data to examine the role of
the $H(z)$ data played in cosmological constraints. Recently, the
Supernova Cosmology Project (SCP) collaboration have released their
Union2 compilation which consists of 557 SNe Ia \citep{Amanullah10},
which have been used to constrain cosmological models in
\citet{Wei10c,Xu10,Liang11a,Liang11c}. The distance modulus of SN Ia
can be given by
\begin{equation}
\mu=5 \log(d_L/\rm{Mpc})+25~,
\end{equation}
where the luminosity distance can be calculated as
$d_L={[c(1+z)/H_0]} \int^{z}_0{dz'}/{E(z')}$. In the calculation of
the likelihood from SNe Ia, we have marginalized the nuisance
parameter \citep{Pietro03}:
\begin{equation}
\label{chi2SN} \chi^2_{\rm
SNe}=A-\frac{B^2}{C}+\ln\left(\frac{C}{2\pi}\right),
\end{equation}
where $A=\sum_i^{557}{(\mu^{\rm data}-\mu^{\rm
th})^2}/{\sigma^2_{\mu,i}}$, $B=\sum_i^{557}{(\mu^{\rm
data}-\mu^{\rm th})}/{\sigma^2_{\mu,i}}$,
$C=\sum_i^{557}{1}/{\sigma^2_{\mu,i}}$, and $\sigma_{\mu,i}$ is the
$1\sigma$ uncertainty of the SNe data.

\section{Constraint on the phenomenological interacting
scenario}\label{sec3}

The model parameters are determined by applying the maximum
likelihood method of $\chi^{2}$ fitting by using the Markov Chain
Monte Carlo (MCMC) method. The total $\chi^2$ with the joint data of
$H(z)$+CMB+BAO+SNe can be given by
\begin{equation} \label{chi2min}
\chi^2=\chi^2_{H}+\chi^2_{\rm BAO}+\chi^2_{\rm CMB}+\chi^2_{\rm
SNe}.
\end{equation}
In adopting the MCMC approach, we generate a chain of sample points
distributed in the parameter space according to the posterior
probability by using the Metropolis-Hastings algorithm with uniform
prior probability distribution, and then repeat this process until
the established convergence accuracy can be satisfied. Our code is
based on CosmoMCMC \citep{Lewis02}.

\begin{figure}
\begin{center}
\includegraphics[width=1.2\hsize]{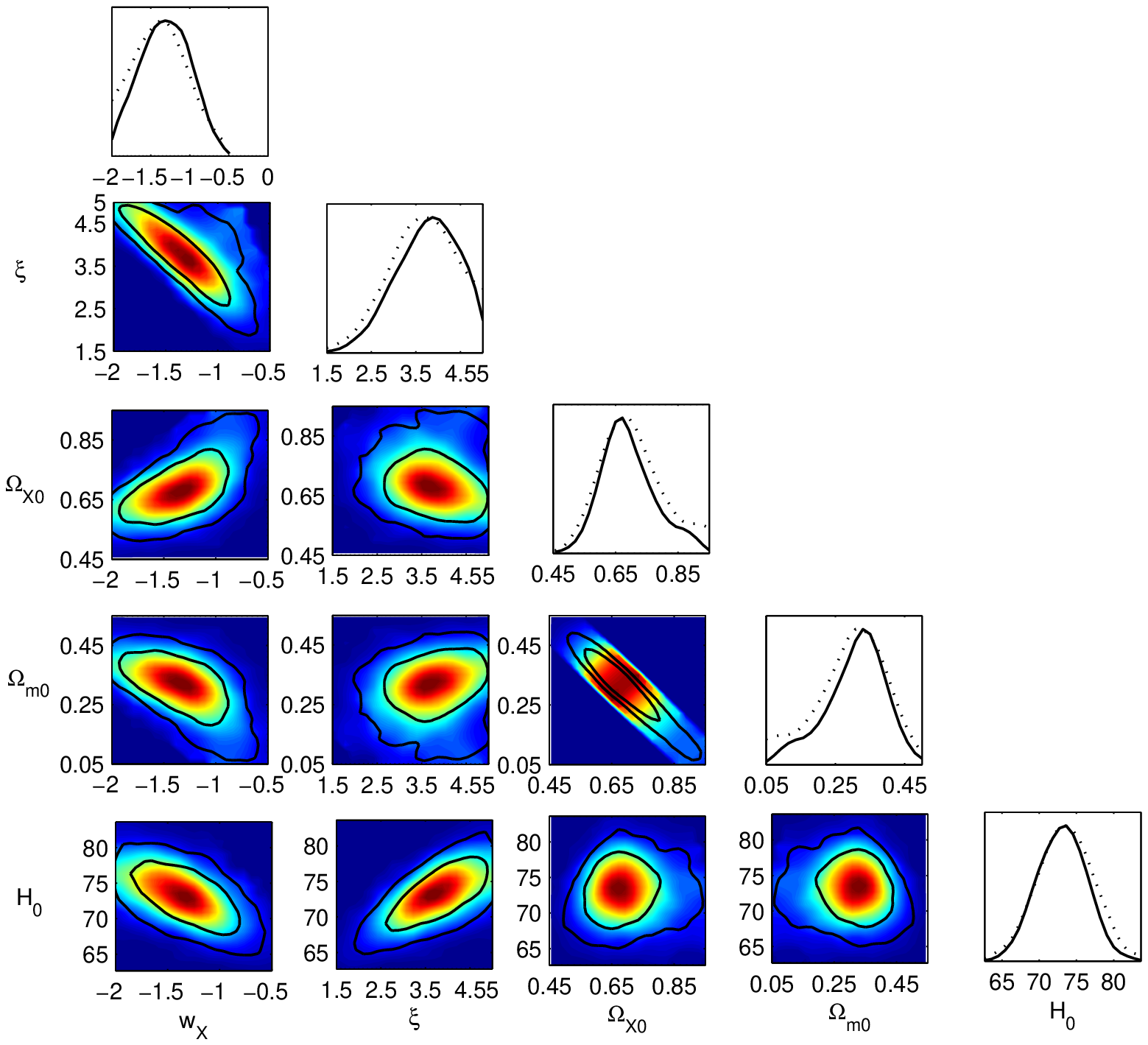}
\end{center}
\caption{The 2-D regions and 1-D marginalized distribution with the
1-$\sigma$ and 2-$\sigma$ contours of parameters
$w_X$, $\xi$, $\Omega_{X0}$, $\Omega_{m0}$,
and $H_0$ in the phenomenological interacting scenario, for the
$H(z)$ data. \label{H}}
\end{figure}

\begin{figure}
\begin{center}
\includegraphics[width=1.2\hsize]{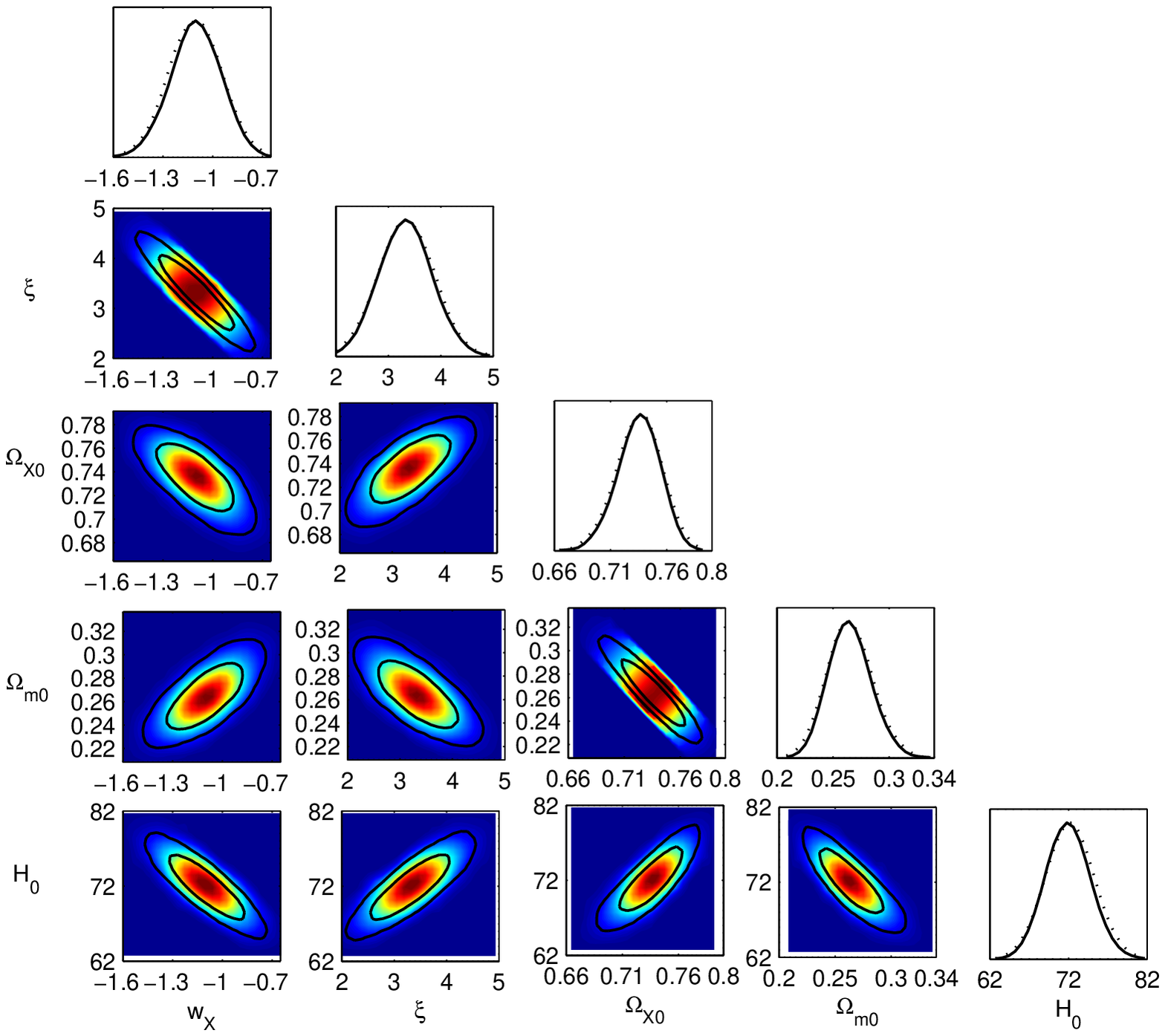}
\end{center}
\caption{The 2-D regions and 1-D marginalized distribution with the
1-$\sigma$ and 2-$\sigma$ contours of parameters
$w_X$, $\xi$, $\Omega_{X0}$, $\Omega_{m0}$,
and $H_0$ in the phenomenological interacting scenario, for the data
sets $H(z)$+BAO+CMB. \label{H+bao+cmb}}
\end{figure}

\begin{figure}
\begin{center}
\includegraphics[width=1.2\hsize]{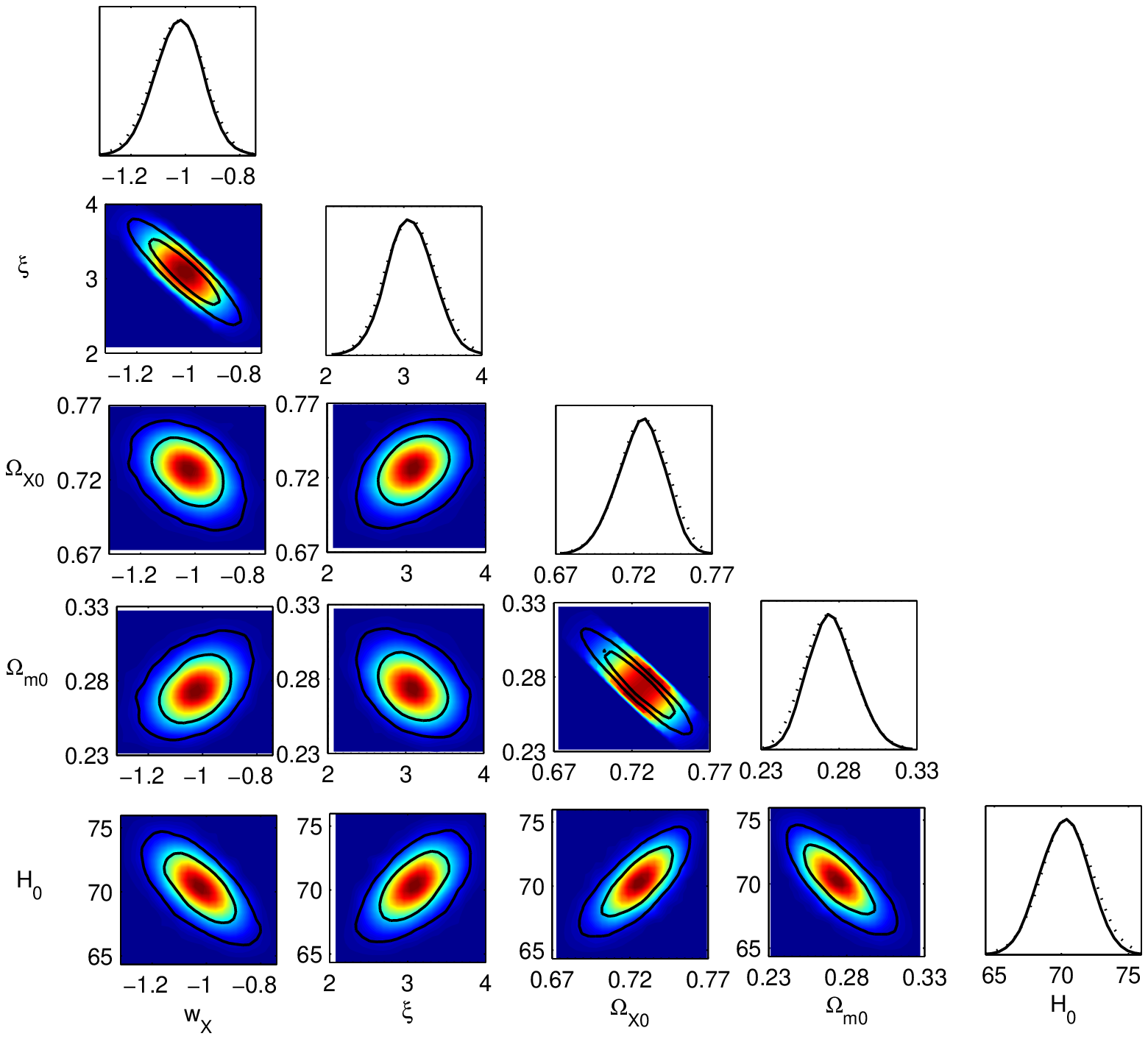}
\end{center}
\caption{The 2-D regions and 1-D marginalized distribution with the
1-$\sigma$ and 2-$\sigma$ contours of parameters
$w_X$, $\xi$, $\Omega_{X0}$, $\Omega_{m0}$,
and $H_0$ in the phenomenological interacting scenario, for the data
sets SNe+BAO+CMB. \label{sn+bao+cmb}}
\end{figure}

\begin{figure}
\begin{center}
\includegraphics[width=1.2\hsize]{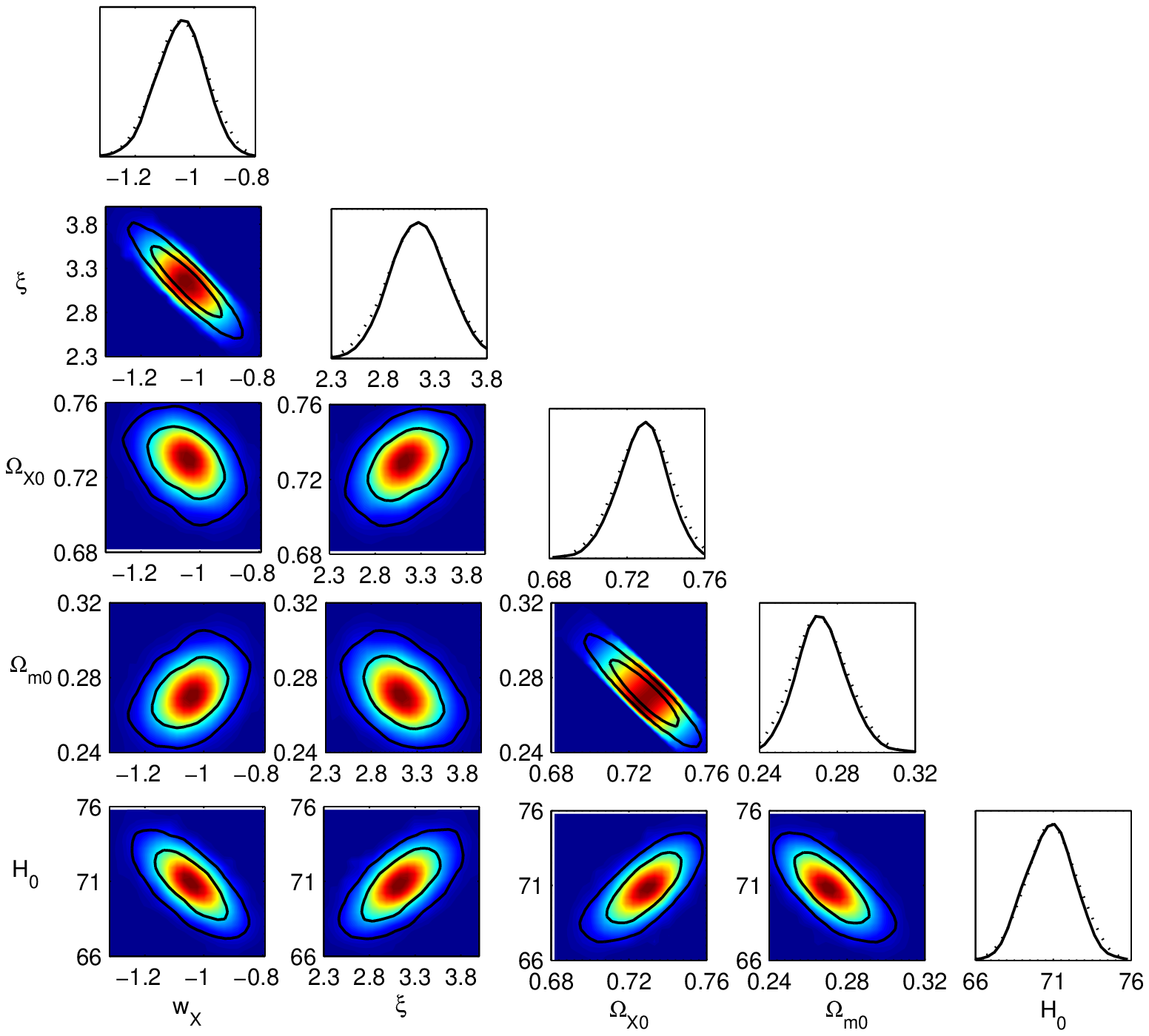}
\end{center}
\caption{The 2-D regions and 1-D marginalized distribution with the
1-$\sigma$ and 2-$\sigma$ contours of parameters
$w_X$, $\xi$, $\Omega_{X0}$, $\Omega_{m0}$,
and $H_0$ in the phenomenological interacting scenario, for the data
sets $H(z)$+SNe+BAO+CMB.\label{all}}
\end{figure}

We show the 1-D probability distribution of each parameter ($w_X$,
$\xi$, $\Omega_{X0}$, $\Omega_{m0}$, $H_0$)  and 2-D plots for
parameters between each other for the phenomenological interacting
scenario with $H(z)$ in Fig.~\ref{H}. The best-fit values of the
model parameters are $\Omega_{m0}=0.32_{-0.19}^{+0.12}$,
$w_X=-1.34_{-0.66}^{+0.58}$ and $\xi=3.72_{-1.42}^{+1.27}$, which
show that the $H(z)$ data only can not tightly constrain the model
parameters. Fitting results from the joint data of $H(z)$+CMB+BAO
are given in Fig~\ref{H+bao+cmb}, with the best-fit values
$\Omega_{m0}=0.26_{-0.03}^{+0.04}$, $w_X=-1.11_{-0.27}^{+0.27}$ and
$\xi=3.35_{-0.92}^{+0.88}$, which show that when combined to CMB+BAO
data, the $H(z)$ data can give more stringent constraints. For
comparison, fitting results from SNe+CMB+BAO without $H(z)$ are
given in Fig~\ref{sn+bao+cmb}, with the best-fit values
$\Omega_{m0}=0.27_{-0.03}^{+0.03}$, $w_X=-1.03_{-0.16}^{+0.16}$ and
$\xi=3.10_{-0.55}^{+0.54}$, which are in good agreement with those
of $H(z)$+BAO+CMB data. The contours constrained with
$H(z)$+SNe+BAO+CMB are shown in Fig.\ref{all}, and the best-fit
values are $\Omega_{m0}=0.27_{-0.02}^{+0.02}$,
$w_X=-1.05_{-0.14}^{+0.15}$ and $\xi=3.15_{-0.50}^{+0.48}$, which
are in good agreement with those of $H(z)$+BAO+CMB and SNe+BAO+CMB.
We present the best-fit values of parameters with 1-$\sigma$ and
2-$\sigma$ uncertainties of the phenomenological interacting
scenario in Table~\ref{result}.

From Fig.~\ref{H}-\ref{all} and Table~\ref{result}, it is shown that
our results are more stringent and consistent with the constraint
results obtained by combing previous SNe Ia data to BAO+CMB data
\citep{Guo07,Chen10,Wei10a}. And the special case ($\xi=3, w_X=-1$,
corresponding to the $\Lambda$CDM with no interaction) is included
at $1\sigma$ confidence level with the recent observational data;
however, it is also shown that the constraints favor
$\xi+3\omega_X>0$ for the phenomenological scenario, and indicate
that the energy is transferred from dark matter to dark energy and
the coincidence problem is quite severe, which is consistent with
those obtained in \citet{Guo07,Chen10}. We also find that the $H(z)$
data can give more stringent constraints on the phenomenological
interacting scenario when combined to CMB and BAO observations.
Comparing the SNe Ia data in the same way, we can find the
confidence regions of $H(z)$+BAO+CMB data are in good agreement with
those of SNeIa+BAO+CMB data; this situation has also been noted in
\citet{Zhai10} for constraining on the $\Lambda$CDM and XCDM
scenario. We also find the confidence regions of $H(z)$+BAO+CMB,
SNe+BAO+CMB, and $H(z)$+SNe+BAO+CMB are consistent with each other.
This situation is similar to that obtained in \citet{Cao11} for
constraining on the Interacting Dark Matter (IDM) scenario with
$H(z)$+SNe data.

\begin{table*}
 \begin{center}{\scriptsize
 \begin{tabular}{|c|c|c|c|c|} \hline\hline
 & \multicolumn{4}{c|}{The phenomenological interacting scenario}  \\
 \cline{1-5}               &$H(z)$    &$H(z)$+BAO+CMB                 &SNe+BAO+CMB                      &$H(z)$+SNe+BAO+CMB          \\ \hline
$w_X$                 \ \ & \ \ $-1.34_{-0.66}^{+0.58}(1\sigma)_{-0.88}^{+0.79}(2\sigma)$ \ \ & \ \ $-1.11_{-0.27}^{+0.27}(1\sigma)_{-0.42}^{+0.38}(2\sigma)$ \ \  & \ \ $-1.03_{-0.16}^{+0.16}(1\sigma)_{-0.24}^{+0.24}(2\sigma)$ \ \ & \ \ $-1.05_{-0.14}^{+0.15}(1\sigma)_{-0.21}^{+0.21}(2\sigma)$\ \ \\
$\xi$              \ \ & \ \ $3.72_{-1.42}^{+1.28}(1\sigma)_{-1.96}^{+1.88}(2\sigma)$ \ \ & \ \ $3.35_{-0.92}^{+0.88}(1\sigma)_{-1.32}^{+1.37}(2\sigma)$ \ \ & \ \ $3.10_{-0.55}^{+0.54}(1\sigma)_{-0.79}^{+0.82}(2\sigma)$\ \  & \ \ $3.15_{-0.50}^{+0.48}(1\sigma)_{-0.71}^{+0.72}(2\sigma)$\ \ \\
$\Omega_{X0}$    \ \ & \ \ $0.68_{-0.12}^{+0.19}(1\sigma)_{-0.17}^{+0.26}(2\sigma)$ \ \ & \ \ $0.74_{-0.04}^{+0.03}(1\sigma)_{-0.06}^{+0.05}(2\sigma)$\ \   & \ \ $0.73_{-0.03}^{+0.03}(1\sigma)_{-0.04}^{+0.04}(2\sigma)$\ \  & \ \ $0.73_{-0.02}^{+0.02}(1\sigma)_{-0.04}^{+0.03}(2\sigma)$\ \ \\
$\Omega_{m0}$     \ \ & \ \ $0.32_{-0.19}^{+0.12}(1\sigma)_{-0.26}^{+0.17}(2\sigma)$ \ \ & \ \ $0.26_{-0.03}^{+0.04}(1\sigma)_{-0.05}^{+0.06}(2\sigma)$\ \   & \ \ $0.27_{-0.03}^{+0.03}(1\sigma)  _{-0.04}^{+0.04}(2\sigma)$\ \    & \ \ $0.27_{-0.02}^{+0.02}(1\sigma)_{-0.03}^{+0.04}(2\sigma)$\ \ \\
$H_0$                 \ \ & \ \ $73.10_{-5.84}^{+6.40}(1\sigma)_{-8.47}^{+8.98}(2\sigma)$ \ \ & \ \ $72.09_{-5.04}^{+5.30}(1\sigma)_{-7.39}^{+8.01}(2\sigma)$\ \            & \ \ $70.44_{-3.25}^{+3.12}(1\sigma)   _{-4.83}^{+4.43}(2\sigma)$\ \      & \ \ $70.96_{-2.84}^{+2.63}(1\sigma) _{-4.08} ^{+4.07}(2\sigma)$\ \ \\
 \hline\hline
$\chi_{\rm min}^2$             \ \ & \ \ $9.851$\ \    \ \ & \ \ $10.935$\ \                  & \ \ $544.117$\ \                   & \ \ $ 555.762$\ \      \\
\hline\hline

 \end{tabular}}
 \end{center}
 \caption{The best-fit values of parameters $w_X$, $\xi$,
 $\Omega_{X0}$, $\Omega_{m0}$, and $H_0$ for the phenomenological scenario
 with the 1-$\sigma$ and 2-$\sigma$ uncertainties, as well as $\chi_{\rm min}^2$, 
for the data sets $H(z)$, $H(z)$+BAO+CMB, SNe+BAO+CMB, and
$H(z)$+SNe+BAO+CMB, respectively. \label{result}}
 \end{table*}

\section{Conclusions}\label{sec4}

In this paper, we test the interacting dark energy scenario with a
phenomenological scaling solution $\rho_X\propto \rho_m a^{\xi}$,
which is proposed as a candidate to ease the coincidence problem of
the concordance $\Lambda$CDM model. With the newly revised
observational $H(z)$ data, the CMB observation from the WMAP7
results, the BAO observation from the SDSS Data Release and the
Union2 SNeIa set, we obtain the best-fit values of the model
parameters in the phenomenological interacting  scenario,
$\Omega_{m0}=0.27_{-0.02}^{+0.02}(1\sigma)_{-0.03}^{+0.04}(2\sigma)$,
$\xi=3.15_{-0.50}^{+0.48}(1\sigma)_{-0.71}^{+0.72}(2\sigma)$, and
$w_X=-1.05_{-0.14}^{+0.15}(1\sigma)_{-0.21}^{+0.21}(2\sigma)$, which
are more stringent and consistent with the previous constraint
results \citep{Guo07,Chen10,Wei10a}.

Our results show that the $\Lambda$CDM model still remains a good
fit to the recent observational data. However, the interaction that
the energy transferring from dark matter to dark energy is slightly
favored over the interaction from dark energy to dark matter, which
is consistent with that obtained in \citet{Guo07,Chen10}, therefore,
the coincidence problem is quite severe in the phenomenological
scenarios. When combined the $H(z)$ data with CMB and BAO
observations, it is shown that the $H(z)$ data can give more
stringent constraints on the phenomenological interacting scenario.
In order to examine the role of the $H(z)$ data played in
cosmological constraints, we compared the SNe Ia data in the same
way and find the constraints with $H(z)$+BAO+CMB, SNeIa+BAO+CMB, and
$H(z)$+SNe+BAO+CMB combinations are consistent with each other. With
a large amount of the observational $H(z)$ data in the future, it is
reasonable to expect that the observational $H(z)$ data will play an
increasingly important role in cosmological researches
\citep{Zhai10,Ma11}.

\section*{Acknowledgments}
We thank Lixin Xu for introducing the powerful program cosmoMCMC and
Yu Pan for some calculations. We are also grateful to Yun Chen, Hao
Wang, Xiaolong Gong, Xin-jiang Zhu, Yan Dai, Fang Huang, Jing Ming,
Kai Liao, Yubo Ma, Huihua Zhao and Dr. Yi Zhang for helpful
discussions. This work was supported by the National Natural Science
Foundation of China under the Distinguished Young Scholar Grant
10825313 and Grant 11073005, the Ministry of Science and Technology
national basic science Program (Project 973) under Grant
No.2007CB815401, the Fundamental Research Funds for the Central
Universities and Scientific Research Foundation of Beijing Normal
University.


\label{lastpage}

\end{document}